\documentclass[twocolumn,prb,superscriptaddress,showpacs,citeautoscript,floatfix]{revtex4}
\usepackage{amsfonts}
\usepackage{amssymb}
\usepackage{amsmath}
\usepackage{graphicx}
\usepackage{dcolumn}   
\usepackage{bm}        
\usepackage{flafter}

\usepackage{hyperref}  

\begin{document}


\title{Structural properties of the sesquicarbide superconductor La$_{2}$C$_{3}$ at high pressure}

\date{\today}

\author{X. Wang}
\author{I. Loa}
\author{K. Syassen}
\email[Corresponding author:~E-mail~]{k.syassen@fkf.mpg.de}
\author{R. K. Kremer}
\author{A. Simon}
\affiliation{Max-Planck-Institut f\"{u}r Festk\"{o}rperforschung,
Heisenbergstrasse 1, D-70569 Stuttgart, Germany}
\author{M. Hanfland}
\affiliation{European Synchrotron Radiation Facility, BP 220,
F-38043 Grenoble, France}
\author{K. Ahn}
\affiliation{Department of Chemistry, Yonsei University, Wonju
220-710, South Korea}

\begin{abstract}
The effect of pressure on the structural properties of lanthanum
sesquicarbide  La$_2$C$_3$ ($T_{\rm c}$ = 13~K) has been investigated
at room temperature by angle-dispersive powder x-ray diffraction in a
diamond anvil cell. The compound remains in the cubic Pu$_2$C$_3$-type
structure at pressures up to at least 30~GPa. The corresponding
equation of state parameters are reported and discussed in terms of
phase stability of La$_2$C$_3$. Pressure-volume data of the impurity
phase LaC$_2$ are reported also for pressures up to 13~GPa.
\end{abstract}

\smallskip

\pacs{PACS: 61.50.Ks, 62.50.+p, 77.80.Bh}

\maketitle

\section{Introduction}

Superconductivity in binary rare earth carbides has been an intensively
investigated topic for many years. Outstanding in this family of
compounds with respect to their superconducting transition temperatures
$T_{\rm c}$ are the sesquicarbides RE$_2$C$_3$, RE being the
nonmagnetic rare earth metals Y and La. These systems recently regained
attention due to the finding by Amano {\it et al.} and Nakane {\it et
al.} who reported the successful synthesis of binary Y$_2$C$_3$ under
high pressure conditions ($\sim$5~GPa). They reported transition
temperatures $T_{\rm c}$ and critical fields $B_{c2}$ of 18~K and
$>$30~T, respectively \cite{Amano04a,Nakan04}.

\begin{figure}[tb]
\centerline{\includegraphics[width=80mm,clip]{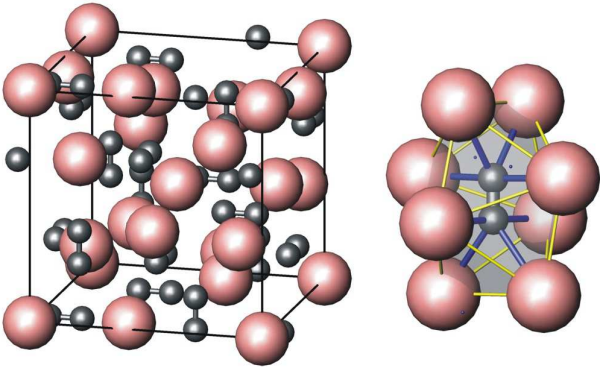}}%
\caption{\label{fig1} (Color online) The Pu$_2$C$_3$-type cubic crystal
structure of La$_2$C$_3$ [space group (SG) $I\overline{4}3d$ (No. 220),
$Z$ = 8 formula units in the cubic cell]. The dicarbide anions occupy
the voids in bisphenoids of the metal substructure. Note: Different
from the standard crystallographic setting, the origin of the indicated
unit cell is shifted into a La position.}
\end{figure}

All RE carbides of composition RE$_2$C$_3$ are isotypic; they
crystallize in the cubic Pu$_2$C$_3$ structure-type \cite{Atoji61} (see
Fig. \ref{fig1}). Electronic structure calculations, based on the
Pu$_2$C$_3$ structure type, have been carried out recently for
Y$_2$C$_3$ by Shein {\it et al.} and by Singh and Mazin
\cite{Shein04,Singh04}. Singh and Mazin \cite{Singh04} specifically
discuss the origin and magnitude of the electron-phonon coupling in
Y$_2$C$_3$. They identify low-frequency metal atom vibrations to have
the largest coupling whereas the contribution of the high-frequency
C--C stretching vibrations is found to be comparatively small.

So far, the samples of Y$_2$C$_3$ showing a $T_{\rm c}$ of 18~K have
been characterized by x-ray powder diffraction only. A more detailed
investigation of these phases with respect to the exact composition and
the phase diagram is still pending. In this respect La$_2$C$_3$, which
shows superconductivity with $T_{\rm c}$ up to 13.2~K and which can be
prepared at ambient pressure, is better understood
\cite{Spedding,Gschneidner,Simon2004}.

We report here the equation of state and the effect of pressure on the
structure of La$_2$C$_{3}$. The structural properties were measured up
to 30~GPa using synchrotron powder x-ray diffraction in the
angle-dispersive mode. Our primary motivation was to investigate how
pressure affects the structural degrees of freedom of the
Pu$_2$C$_3$-type phase; these are thought to play a key role in
controlling the details of the electronic structure in the vicinity of
the Fermi level. Second, the equation of state and phase stability of
La$_2$C$_3$ is of interest in the context of pressure synthesis of rare
earth carbides. To our knowledge, pressure-dependent studies similar to
the ones presented here have not been reported so far for any of the
known rare earth and actinide sesquicarbides.

\begin{figure}[tb]
\centerline{\includegraphics[width=70mm,clip]{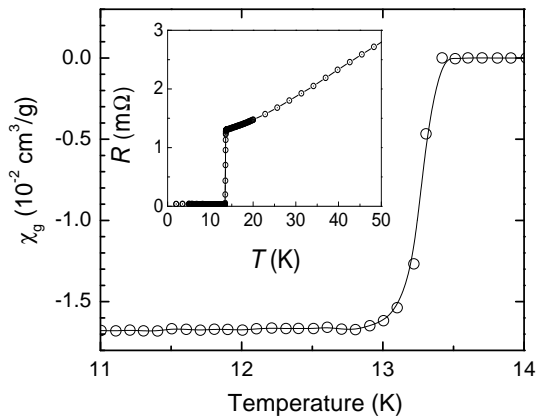}}
\caption{\label{suscep} Diamagnetic shielding and electrical resistance
(inset) of the La$_2$C$_3$ sample used for the high-pressure x-ray
diffraction experiment. The diamagnetic shielding was measured in a
field of 0.7~mT after the sample had been cooled in zero field to 2~K.}
\end{figure}

\section{Sample characterization, diffraction experiments}

The superconducting properties of La$_2$C$_3$ depend sensitively on the
carbon content. Spedding {\it et al.} and Gschneidner {\it et al.}
report a homogeneity regime for La$_2$C$_{3-x}$ which ranges from 56.2
at.-\% to 60.2 at.-\% carbon with cubic lattice parameters from 8.803
\AA~ to 8.818 \AA, respectively \cite{Spedding,Gschneidner}. Recently,
Simon and Gulden reinvestigated the phase diagram of La$_2$C$_{3-x}$
($0<x<0.33$) \cite{Simon2004}. They find that the boundary phases with
$x=0$ and $x=0.33$ are homogeneous but that by sufficiently extended
annealing, samples within the homogeneity range tend to phase separate
into the boundary phases with $x=0$ and $x=0.33$. According to magnetic
susceptibility measurements these boundary phases have sharp
superconducting transitions at 5.6~K and at 13.2~K, respectively
\cite{GuldenThesis}. The latter value obtained for the stoichiometric
phase La$_2$C$_3$ is appreciably increased over the early $T_{\rm c}$
of 11~K reported by Giorgi {\it et al} \cite{Giorgi1969,Giorgi1970}.
The 13.2~K phase shows critical fields $B_{ c2} \geq 17$~T,
significantly enhanced over $B_{ c2} = 12$~T found by Francavilla {\it
et al.} \cite{Francavilla} (see also
Ref.\mbox{~}\onlinecite{HennThesis}).

The polycrystalline sample of La$_2$C$_3$ used in the present study was
prepared by arc melting appropriate quantities of La metal (Alfa,
99.99\%) and graphite chips (Deutsche Carbone, 99.99\%) with a slight
excess (2 to 3\%) of graphite. The graphite chips had been heated
(1050~$^{\circ}$C) and degassed in vacuum ($10^{-5}$\,mbar) for one day
and then stored in a dried argon atmosphere. The sample pellets were
annealed at 650~$^{\circ}$C for 3 days and cooled to room temperature
at a rate of 10 $^{\circ}$C/hr. All subsequent sample manipulations,
including preparations for pressure experiments, were done in dry argon
atmosphere.

\begin{figure}[tb]
\centerline{\includegraphics[width=70mm,clip]{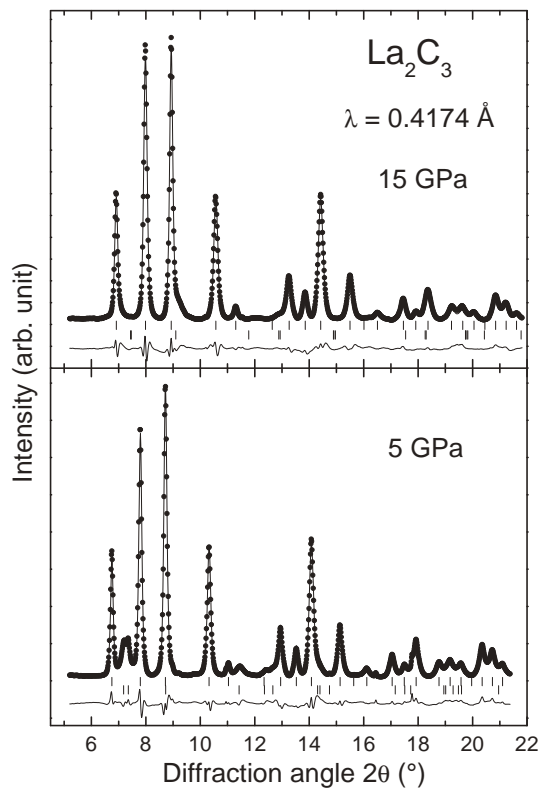}}
\caption{\label{fig3} Observed, calculated, and difference x-ray powder
diffraction patterns for La$_2$C$_3$ at 5.0 and 15~GPa. The diffraction
patterns were refined assuming an admixture of a tetragonal LaC$_2$
phase \cite{Jones91}. Vertical markers indicate Bragg reflections of
the two phases.}
\end{figure}

The phase composition was checked by laboratory-based powder x-ray
diffraction measurements using Cu\,K$\alpha_1$ radiation; these showed
La$_2$C$_3$ as the majority phase ($\sim$85\,\%) with a lattice
parameter of 8.8150(6)~\AA. Weak additional reflections were observed
which are attributed to an impurity phase of tetragonal LaC$_2$
($a$=3.9323(6)~\AA, c=6.574(1)~\AA), its overall fraction being less
than $\sim$15\,\%. The superconducting properties of the sample were
determined by resistivity, magnetic susceptibility, and heat capacity
measurements. The resistive transition (midpoint) and the onset of
diamagnetic shielding occurs at $T_c^{\rm onset}$=13.5(1)~K (see Fig.
\ref{suscep}). The width of the resistive transition (10\%-90\%
criterion) amounts to 0.1~K. The heat capacity proves bulk
superconductivity  with a characteristic anomaly at $T_{\rm c}$ of
$\Delta C_p/(\gamma T_c) \approx 2.2$; the deviation from the standard
BCS anomaly is most likely due to enhanced electron-phonon coupling.

For the synchrotron x-ray diffraction measurements the sample was
ground to a fine powder and a small amount of the powder was
transferred into the gasket of a diamond anvil cell. Nitrogen was used
as a pressure medium. When nitrogen solidifies (2.5~GPa at 300~K), it
causes some additional Bragg reflections (typically weak) due to its
various high-pressure phases. Angle-dispersive powder x-ray diffraction
patterns (wavelength $\lambda = 0.4176$~\AA) were measured at the
beamline ID9 of the European Synchrotron Radiation Facility, Grenoble,
using image plate detection. The images were integrated using the
program FIT2D to yield intensity versus 2$\theta$ diagrams
\cite{Hamme96}. To improve powder averaging, the DAC was oscillated by
$\pm$3$^\circ$. The ruby luminescence method was used for pressure
measurement \cite{Pierm75,Mao86}.

\section{Results and Discussion}

Figure \ref{fig3} shows representative diffraction diagrams of
La$_2$C$_3$ collected at 5 and 15~GPa. The cubic phase of La$_2$C$_3$
was observed up to the highest pressure of this study, 30.7~GPa, and
after releasing pressure. Extra Bragg reflections seen in the
diffraction diagrams are consistent with the presence of tetragonal
LaC$_2$ (SG $I4/mmm$, No. 139, $Z$=4)e. The reflections due to LaC$_2$
gradually disappeared in the pressure range 10-13~GPa. Upon releasing
pressure, the tetragonal phase was clearly present again at ambient
conditions.

The diffraction patterns were analyzed by the Rietveld method using the
program GSAS \cite{GSAS,Toby01}. The refined parameters of the
sesquicarbide phase (space group $I\overline{4}3d$) were the lattice
constant, the fractional coordinate ($x,x,x$) of the 16c (La) site, a
common isotropic thermal parameter for all atom sites, a Chebyshev
polynomial background, Pseudo-Voigt profile function parameters, and an
overall intensity scaling factor. The refinements were not sensitive to
the exact value of the positional parameter $v,0,0.25$ of the carbon
24d site. Hence, $v$ was fixed at 0.3049
(Ref.\mbox{~}\onlinecite{Atoji61}), corresponding to a C--C distance of
1.236~\AA~ at ambient, and assumed independent of pressure. A preferred
orientation correction was applied, but was found to result in only
minor improvements of the refinements. Two-phase refinements were
performed in order to account for the admixture of tetragonal LaC$_2$,
but only the unit cell parameters were refined for this phase; its
internal parameter for the carbon position was assumed constant
($z$=0.4024 \cite{Jones91}, corresponding to a C--C distance of
1.284~\AA~ at ambient). Figure~\ref{fig3} illustrates the results of
the refinements for the two patterns collected at 5.0 and 15.0~GPa. The
convergence was achieved at residuals (with a subtracted background)
R$_{wp}$ = 8.7\% for 5~GPa and at R$_{wp}$ = 8.2\% for 15~GPa.

\begin{figure}[tb]
\centerline{\includegraphics[width=70mm,clip]{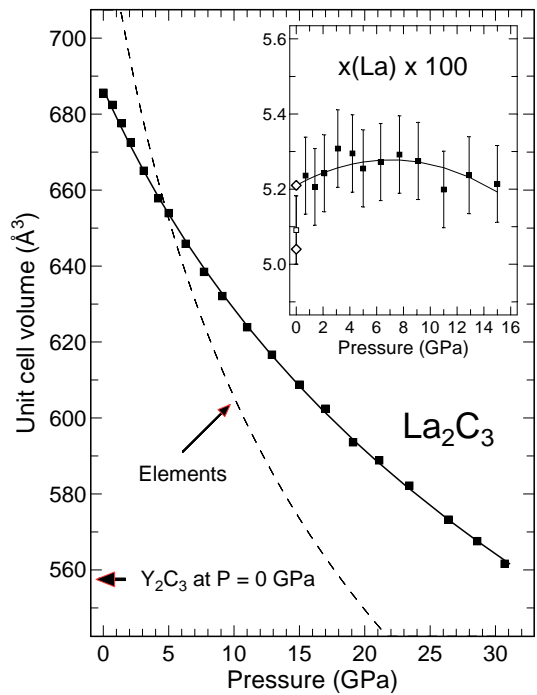}}
\caption{\label{fig5} Unit cell volume of La$_2$C$_3$ as a function of
pressure. The solid line refers to a fitted Birch relation. The dashed
line sketches the $PV$ relation of the elemental constituents La metal
and carbon (diamond form) taken in appropriate amounts. The inset shows
the positional parameter $x$(La) as a function of pressure. Solid
symbols are for data measured under pressure, open symbols are used to
represent the $x$-value obtained after pressure cycling and
corresponding literature data \cite{Atoji61,Simon2004}.}
\end{figure}

The experimental pressure-volume data of La$_2$C$_3$ are shown in
Fig.~\ref{fig5}. At 30~GPa, La$_2$C$_3$ is compressed by 18\% and the
volume is comparable to that of Y$_2$C$_3$ at ambient conditions. The
pressure-volume data were fitted by a Birch equation of state
\cite{Birch78}
\begin{equation}\label{Birch}
P(x) = \frac{3}{2} B_0 \cdot [x^{-7} - x^{-5}]\\ \times\, \left[1 -
\frac{3}{4} (B_0^{\prime}-4)(1-x^{-2})\right],
\end{equation}
where $x = (V/V_0)^{1/3}$ is a reduced length. The parameters are the
volume $V_0$, the bulk modulus $B_0$, and its pressure derivative
$B_0^{\prime}$, all at zero pressure. The fitted parameters and their
standard deviations are given in Table \ref{tab3}. The obtained value
for $V_0$ is consistent with literature data \cite{Atoji61,Simon2004}.

\begin{table}[tb]
\caption{\label{tab3} Summary of fitted equation-of-state parameters
($V_0$, $B_0$, and $B^{\prime}$, see Eq. \ref{Birch}) for cubic
La$_2$C$_3$ and tetragonal LaC$_{2}$. The $V_0$ values refer to the
volume of the respective conventional unit cell. Also listed is
$V_0^{\rm meas}$ which refers to the ambient-pressure value measured
with Cu\,K$\alpha_1$ radiation.}
\bigskip%
\begin{ruledtabular}
\begin{tabular}{lllll}
Compound & $V_0^{\rm meas}$ (\AA$^3$) & $V_0$ (\AA$^3$) & $B_0$ (GPa)& $B^{\prime}$\\
\hline
La$_2$C$_3$ & 684.96& 686.6(4) & 89.0(20) & 5.5(3) \\
LaC$_2$    &  101.65 & 102.3(3) & 76.3(90) & 4.6(15) \\
\end{tabular}
\end{ruledtabular}
\end{table}

Among the sesquicarbide compounds with the cubic Pu$_2$C$_{3}$-type
structure, La$_2$C$_{3}$ has the largest lattice parameter and
therefore is expected to represent the most compressible candidate in
this family of compounds. There appears to be no experimental data
available on the bulk moduli of other cubic rare earth sesquicarbides
to compare with. The bulk modulus of La$_2$C$_{3}$ at 30~GPa is about
250~GPa. This value would be an upper limit for the ambient-pressure
bulk modulus $B_0$ of Y$_2$C$_3$; it is much lower than the calculated
one ($B_0$ = 363 GPa) reported in Ref.\mbox{~}\onlinecite{Shein04}.

At ambient conditions, the volume per La$_2$C$_3$ formula unit (85.68
\AA$^3$) is smaller than that of the constituents La metal and carbon
in the diamond modification (La metal 37.5 \AA$^3$/La, diamond 5.67
\AA$^3$/C, 2 $\times$ 37.5 + 3 $\times$ 5.67 = 92.0 \AA$^3$). So, the
$P\Delta V$ term in the free energy difference favors the formation of
the compound under pressure. However, because of the large
compressibility of La metal [$B_0$(La) $\approx$ 25~GPa,
Ref.\mbox{~}\onlinecite{Syass75a}], the $P\Delta V$ term changes sign
at about 5~GPa (Fig.~\ref{fig5}). Hence, La$_2$C$_3$ is not necessarily
a thermodynamically stable phase at all pressures covered in our
experiment. A high-temperature experiment performed above 5~GPa may
result in decomposition or, perhaps, the formation of a metal-rich
carbide.

The Pu$_2$C$_3$-type structure is closely related to the
anti-Th$_3$P$_4$-type. This similarity becomes obvious when writing the
stoichiometry as La$_4$(C$_2$)$_3$ instead of La$_2$C$_3$. The metal
atoms in the Pu$_2$C$_3$ structure form a 3D network of condensed
RE$_8$ bisphenoids. The dicarbide anions are centered at the
interstitial 12a site ($3/8,\,0,\,1/4$) and oriented along the
$\overline{4}$ axes of the bisphenoids (Fig. \ref{fig1}) whose shape
depends on the positional parameter $x$ \cite{Hyde89,Carri99}. For
La$_2$C$_3$ we have $x$(La) $\approx 0.05$ which is intermediate
between nearly equal lengths of all edges ($x \approx 1/32$) and equal
center-corner distances ($x=1/12$). Near $x=0.05$ the metal-metal
coordination is $3+2+6$ and the bisphenoids can host elongated
entities.

Within the uncertainty of our data, the positional parameter $x$(La)
stays constant under pressure (Fig. \ref{fig5}). This means that the
coordination polyhedron around C$_2$ is compressed isotropically, and
there is no change in the second neighbor coordination. Actually, at
ambient pressure the $x$ parameters of all cubic rare earth and
actinide sequicarbides cluster around 0.051 with a quite narrow spread
of $\pm 0.002$ \cite{ICSD}.

The arrangement of metal ions in La$_2$C$_3$ is similar to that of the
'cI16' (16-atom body centered cubic) phase of high-pressure Li which
can be interpreted as a $2 \times 2 \times 2$ superstructure of bcc
\cite{Hanfl00}. In Li-cI16 a pseudo-gap is formed, involving a lowering
of the density of states $N(E_F)$ at the Fermi level, which leads to a
decrease of the total energy relative to any of the common
high-symmetry phases of elemental metals. Despite the pseudo-gap
formation, the cI16 phase of Li is a superconductor with $T_{\rm c}>$
10~K (see Ref.\mbox{~}\onlinecite{Deemy03} and literature cited
therein). As for La$_2$C$_3$, the Fermi level is calculated to also
fall into a local minimum of the DOS \cite{Kim05}. This is
qualitatively similar to what is reported for Y$_2$C$_3$
\cite{Shein04,Singh04}. It is not clear whether the formation of a
local DOS minimum in the sesquicarbides is a consequence of the
structural distortion of the metal sublattice away from bcc. In any
case, the states in the DOS minimum of Y$_2$C$_3$ show a strong
coupling to 'symmetry-preserving' metal atom displacements, i.e. a
change in the $x$ parameter \cite{Singh04}.

\begin{figure}[tb]
\centerline{\includegraphics[width=55mm,clip]{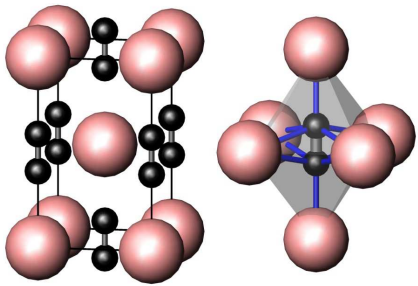}}%
\caption{(Color online) \label{LaC2str} The tetragonal CaC$_2$-type
crystal structure of LaC$_2$ [SG $I4/mmm$ (No. 139), $Z$ = 2,
$a$=3.937, $c$=6.580, La in 2a (0,0,0), C in 4e (0,0,$z$), $z$=0.4026,
distance C--C = 1.289 \AA, after Ref.
\mbox{~}\protect\onlinecite{Jones91}]. The C$_2$ units, oriented along
$c$, are sixfold coordinated by La.}
\end{figure}

For La$_2$C$_3$, a C--C distance of 1.236~\AA~ is given in
Ref.\mbox{~}\onlinecite{Atoji61}. A slightly larger distance, about
1.29~\AA~ (Ref.\mbox{~}\onlinecite{Simon2004}), appears more plausible.
Unfortunately, our experiments do not provide any information on the
pressure dependence of the C--C distance. This distance will correlate
with the degree of charge transfer between La $d$ and antibonding
($\pi^*$) states of the C$_2$ unit. We can expect that the
back-donation effect increases with pressure.

\begin{figure}[tb]
\centerline{\includegraphics[width=70mm,clip]{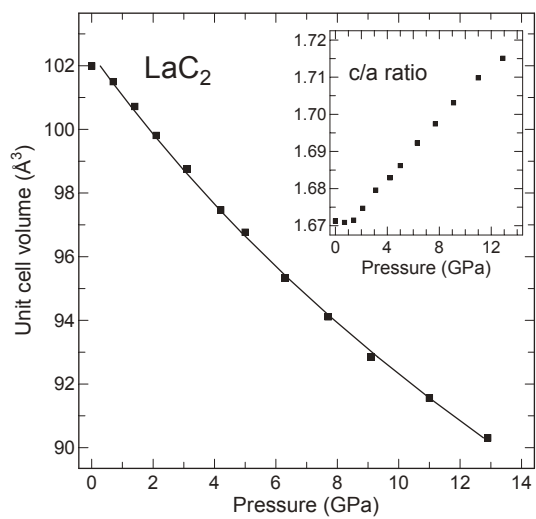}}%
\caption{\label{fig11} Unit cell volume of tetragonal LaC$_2$ as a
function of pressure. The solid line refers to a fitted Birch relation.
The inset shows the $c/a$ ratio as a function of pressure.}
\end{figure}

We now turn to LaC$_2$, the impurity phase in our sample. LaC$_2$
crystallizes in a body-centered tetragonal structure
(Fig.~\ref{LaC2str}). The C$_2$ units in LaC$_2$ (C--C distance 1.28
\AA) are reported to be oriented parallel to the $c$ axis
\cite{Bowma68,Jones91}; they are sixfold coordinated by La atoms which
form elongated octahedra.

The pressure-volume data of LaC$_2$ are shown in Fig. \ref{fig11}.
Parameters obtained by fitting a Birch relation to the high-pressure
data are given in Table \ref{tab3}. The back-extrapolated zero pressure
volume given in Table \ref{tab3} is 0.3\% larger than $V_0 =
102.0$~\AA$^3$ measured after pressure cycling. The latter value agrees
with Ref.\mbox{~}\onlinecite{Jones91}. Despite the larger carbon to
metal ratio, the bulk modulus of LaC$_2$ comes out slightly lower
compared to La$_2$C$_3$. This may be explained by the larger
metal-metal distance in LaC$_2$ (12 $\times$ 3.93 \AA~ at ambient
pressure) compared to the average value for La$_2$C$_3$ (3 $\times$
3.63~\AA, 2 $\times$ 3.81~\AA, 6 $\times$ 4.01~\AA).  The inset of Fig.
\ref{fig11} illustrates that the $c/a$ ratio of LaC$_2$ increases with
increasing pressure, i.e., the compressibility is smaller along the
direction of the dumbbell orientation.

The diffraction pattern of LaC$_2$ was completely lost at pressures
above 13~GPa, but it reappeared upon releasing pressure. It is left to
a separate study to find out whether LaC$_2$ undergoes a reversible
pressure-induced phase transition. Already at ambient pressure, the
volume of LaC$_2$ is larger than that of the constituents (LaC$_2$ = 51
\AA$^3$ per formula unit at $P$=0, La + 2 C = 48.84 \AA$^3$). So,
application of pressure should be a strong driving force for a phase
change, not ruling out at this point a pressure-induced amorphization
as a precursor to phase separation. Another question, brought up by
studies of CaC$_2$ at ambient pressure \cite{Long92,Knapp01}, concerns
the orientation of the C$_2$ dumbbells in tetragonal LaC$_2$. Our data
point to a possible $c/a$ anomaly near ambient pressure (see inset of
Fig. \ref{fig11}) which could be related to a pressure-dependent
reorientation.

In conclusion, the present high-pressure structural study gives
quantitative information on the equation of state and structural
parameters of superconducting La$_2$C$_3$. The results may serve as a
reference for the as yet unknown high-pressure behavior of other cubic
rare earth sesquicarbides. The results are also believed to provide
useful input for modelling the pressure-dependence of the electronic
structure and the electron-phonon coupling of La$_2$C$_3$.





\end{document}